\documentstyle[aps,prl,floats,epsf]{revtex}

\begin{document}

\draft

\wideabs{

%\preprint{ }

\title{Measurement of Low-Energy Cosmic-Ray Antiprotons at Solar Minimum}
\author{
 H.~Matsunaga,$^{1}$
 S.~Orito,$^{1,}$\cite{Oriton}
 H.~Matsumoto,$^{2}$
 K.~Yoshimura,$^{1}$
 A.~Moiseev,$^{3}$
 K.~Anraku,$^{1}$
 R.~Golden,$^{4,}$\cite{Golden}
 M.~Imori,$^{1}$
 Y.~Makida,$^{5}$
 J.~Mitchell,$^{3}$
 M.~Motoki,$^{2}$
 J.~Nishimura,$^{6}$
 M.~Nozaki,$^{2}$
 J.~Ormes,$^{3}$
 T.~Saeki,$^{1}$
 T.~Sanuki,$^{1}$
 R.~Streitmatter,$^{3}$
 J.~Suzuki,$^{5}$
 K.~Tanaka,$^{5}$
 I.~Ueda,$^{1}$
 N.~Yajima,$^{6}$
 T.~Yamagami,$^{6}$
 A.~Yamamoto,$^{5}$
 and
 T.~Yoshida$^{5}$\\
% \begin{center} (BESS Collaboration) \end{center}
}
\address{
$^{1}$University of Tokyo, Tokyo, 113-0033, Japan\\
$^{2}$Kobe University, Kobe, Hyogo, 657-8501, Japan\\
$^{3}$National Aeronautics and Space Administration,
 Goddard Space Flight Center (NASA/GSFC), Greenbelt, MD, 20771\\
$^{4}$New Mexico State University, Las Cruces, NM, 88003\\
$^{5}$High Energy Accelerator Research Organization (KEK),
 Tsukuba, Ibaraki, 305-0801, Japan\\
$^{6}$The Institute of Space and Astronautical Science (ISAS), Sagamihara,
 Kanagawa, 229-8510, Japan\\
}
\date{\today}
\maketitle

\begin{abstract}
The absolute fluxes of the cosmic-ray antiprotons at solar minimum
 are measured in the energy range 0.18 to 1.4 GeV,
 based on 43 events unambiguously detected in BESS '95 data.
The resultant energy spectrum appears to be flat below 1 GeV,
 compatible with a possible admixture of primary antiproton component
 with a soft energy spectrum,
 while the possibility of secondary antiprotons alone explaining the data
 cannot be excluded with the present accuracy.
Further improvement of statistical accuracy and extension of the energy range
 are planned in future BESS flights.
\end{abstract}


}

\narrowtext
The strict upper limit \cite{SA98} on the cosmic-ray antihelium/helium ratio
 provides most direct evidence for Galaxy and nearby part of Universe
 being composed solely of particles.
If so, antiproton ($\bar{p}$) in cosmic-ray must have been produced
 in pairs with protons ($p$'s) by some elementary-particle processes
 in Galaxy or in Universe.
One such process which should certainly exist is
 the interaction of high energy cosmic rays with interstellar gas.
The energy spectrum of $\bar{p}$'s from this ``secondary'' process is expected
 to show a characteristic peak at $2 \sim 3$ GeV
 and sharp decreases of the flux below and above the peak.
One can also conceive more novel or exotic elementary-particle processes
 such as the annihilation of neutralino dark matter
 or the evaporation of primordial black hole \cite{HA75}.
The $\bar{p}$'s from these ``primary'' sources
 are expected to show very soft energy spectra \cite{MA96},
 peaking toward the lower energies,
 and would exhibit large solar modulations \cite{MI96}.

Although general characteristics of cosmic-ray $\bar{p}$ data
 accumulated over last 20 years seem compatible with the secondary $\bar{p}$,
 poor statistics of the data do not allow further investigation.
A detector with much larger acceptance is needed
 both to determine the secondary $\bar{p}$ flux
 and to perform a sensitive search for low-energy primary $\bar{p}$ component.
We report here (for detail see \cite{MATSU}) on such an attempt
 with the BESS spectrometer,
 which was designed \cite{OR87} and constructed for this purpose.
By using BESS '93 data, the first mass-identified detection
 of cosmic-ray $\bar{p}$'s was reported \cite{YO95,MO97}
 based on four events detected
 in the low-energy range of 0.3 to 0.5 GeV.
This measurement (BESS '95) extends the range to 1.4 GeV,
 and was conducted at a period close to solar minimum.

\begin{figure}[b]
\centerline{\epsfxsize=8cm \epsffile{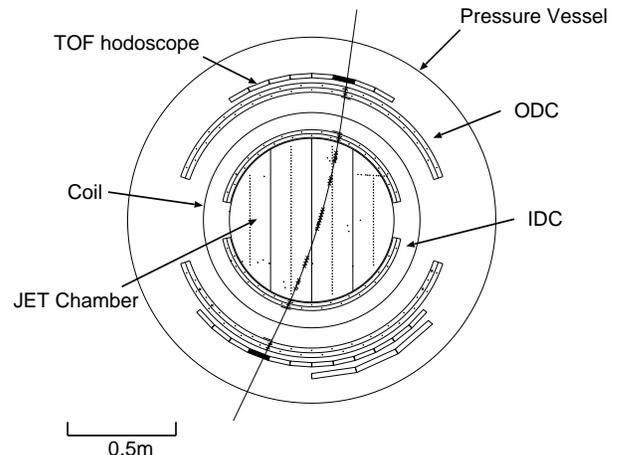}}
\caption{
Cross-sectional view of the BESS '95 detector with one of the $\bar{p}$ events.
}
\label{fig:detector}
\end{figure}

The BESS detector \cite{YOSHIDA} is shown in Fig.\ref{fig:detector}.
The thin superconducting coil \cite{YA88}
 (4 g/cm$^2$ thick including the cryostat)
 produces a uniform axial magnetic field of 1 Tesla.
The $r\phi$-tracking in the central region is performed
 by fitting up to 28 hit-points, each with 200 $\mu$m resolution,
 in the JET- and IDC-drift chambers,
 resulting in a magnetic-rigidity \cite{RIGIDITY} resolution
 of 0.5 \% at 1 GV/$c$.
Tracking in the $z$-coordinate is done to an accuracy of 300 $\mu$m
 by fitting points in IDC measured with vernier pads
 and points in the JET chamber measured using charge-division.
The continuous and redundant 3-dimensional tracking with the drift chambers,
 all equipped with multi-hit capacity, enables us to recognize
 multi-track events and tracks having interactions and scatterings,
 thus minimizing the background.
The $dE/dx$ of the particle in the JET chamber is obtained
 as a truncated mean of the integrated charges of the hit-pulses
 which compose the track.
The TOF scintillator hodoscopes, newly built for '95 flight,
 measure the time-of-flight of particles with a resolution of 110 ps
 as compared to previous 280 ps.

The first-level trigger is provided
 by a coincidence between the top and the bottom scintillators,
 with the threshold set at 1/3 of the pulse height
 from vertically incident minimum ionizing particles.
The second-level trigger,
 which utilizes the IDC and ODC hit-patterns,
 first rejects the null- and multi-track events (pattern-selection)
 and makes a rough rigidity-determination
 to select negatively-charged particles (rigidity-selection).
In addition, one of every 90 first-level triggers is recorded,
 irrespective of the second-level trigger condition,
 to build a sample of unbiased triggers
 from which the efficiencies are determined.
The '95 BESS flight was carried out on July 25, from Lynn Lake, Canada.
The scientific data were taken at an altitude of 36 km
 (a residual atmosphere of 5 g/cm$^2$)
 with a rigidity cutoff ranging 0.3 to 0.5 GV/$c$.
During the live time of $2.72 \times 10^4$ sec,
 4.6 million events were recorded on magnetic tapes,
 out of 60 million cosmic rays which passed through the detector.

The following off-line selections are applied equally
 for positive- and negative-rigidity events.
(i) Only one counter is hit in each layer of the TOF hodoscopes;
(ii) Only one track, which is fully contained in the fiducial region,
 should be found in the JET chamber.
The following cuts are then applied
 to ensure the qualities of the track and the timing measurements.
(1) The fitted $r\phi$- and $z$-track, respectively, should contain
 at least 10 and 5 hits in the JET chamber,
 and at least one hit in each of the upper two and lower two IDC layers;
(2) The reduced $\chi^{2}$ of both the fitted $r\phi$- and $z$-track
 have to be less than 5;
(3) The extrapolated track should cross
 the fiducial region of TOF scintillators ($|z|<46.5$ cm);
(4) The $z$-position determined by the left-right time difference
 measured by the PMTs should match the $z$-impact point
 of the extrapolated track at the TOF counter within 3.8 cm;
(5) The ratio of the signal amplitude of the left and right PMTs
 should be consistent with the $z$-impact point of the extrapolated track.
The combined efficiency of these off-line selections
 is found to be 81 \% at 0.7 GV/$c$ and 78 \% at 2 GV/$c$ for the protons.

\begin{figure}[b]
\centerline{\epsfxsize=8cm \epsffile{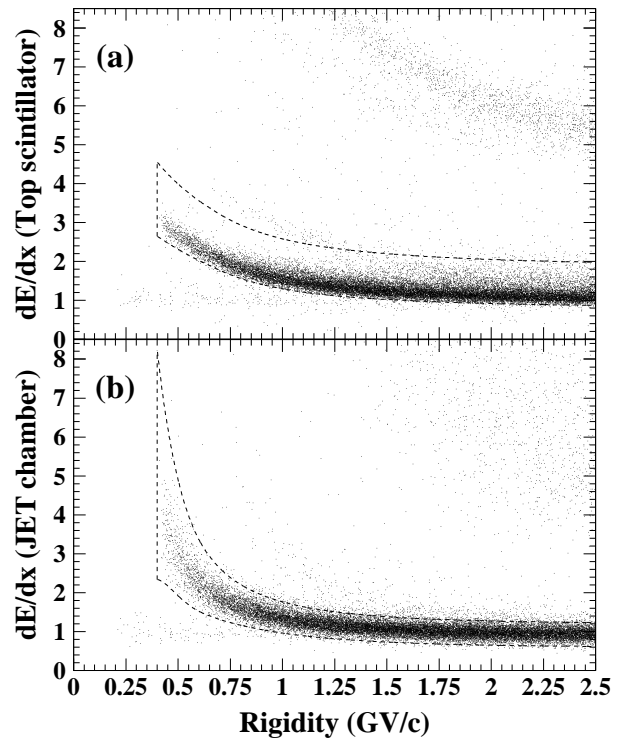}}
\caption{
Scatter plot of $dE/dx$ vs. rigidity
 (a) for the top scintillator and (b) for the JET chamber.
The dashed lines indicate
 the $dE/dx$ band for the proton/antiproton selection.
}
\label{fig:dedxvsr}
\end{figure}

\begin{figure}[t]
\centerline{\epsfxsize=8.11cm \epsffile{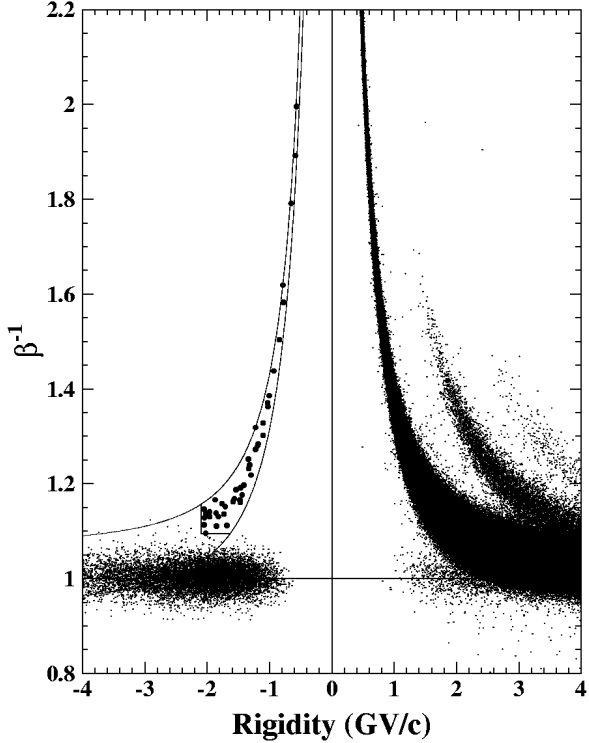}}
\caption{
The identification of $\bar{p}$ events.
The solid lines define the $\beta^{-1}$--$R$ region
 and the $\bar{p}$ mass band used for the spectrum measurement.
}
\label{fig:bvsr}
\end{figure}

We then require that
 the $\bar{p}$'s and the protons must have proper $dE/dx$
 in the top and bottom scintillators as well as in the JET chamber
 (shown in Fig.\ref{fig:dedxvsr}).
These $dE/dx$ selections reject
 all of the particles with charge greater than 1
 and most of the low-energy $e/\mu$,
 while keeping 90 \% efficiency for protons.
The $\beta^{-1}$ vs. rigidity($R$) plot for the remaining events
 is shown in Fig.\ref{fig:bvsr}.
A clear band of $\bar{p}$'s is visible
 at the position exactly symmetric to the proton band.
Each of the $\bar{p}$ candidates was closely investigated
 on the event display for correctness of the track fitting etc.,
 and no particular problems were found.
The probability of positive-rigidity particles or albedo (up-going) particles
 faking any of the $\bar{p}$ candidates is negligibly small.
Also it is apparent from the cleanness of the $\bar{p}$-band
 that there is no background, particularly in the high $\beta^{-1}$ region.
We limit further analysis to the region $\beta^{-1}>1.095$
 in order to have a negligible contamination (less than 0.5 events)
 due to the tail of the $e^{-}/\mu^{-}$ distribution.
The rigidity region is limited to $|R|<2.1$ GV/$c$,
 since the efficiency of the above $\beta^{-1}$ cut becomes lower than 50 \%
 outside of this $|R|$-range.
As seen in Fig.\ref{fig:bvsr},
 all of the 43 negatively-charged events
 observed in this $\beta^{-1}$--$R$ region
 reside inside the 97 \% inclusion band expected for the $\bar{p}$'s.
Furthermore,
 the masses determined for these events show the correct central value,
 and the distribution is similar to the one observed for the protons.
All these background-studies and the consistency-checks allow us to conclude
 that the 43 events are real $\bar{p}$'s incident on the BESS spectrometer.
To check against the possibility that
 any of the $\bar{p}$'s were re-entrant albedo,
 we numerically traced back the trajectories of all $\bar{p}$'s
 through the Earth's geomagnetic field
 and confirmed that all events came from outside of geomagnetic sphere.

\widetext
\begin{table}[b]
%\squeezetable
\caption{Summary of antiproton flux measurement.}
\label{tab:pbsummary}
\begin{tabular}{ccccccccccc}
TOA Kinetic energy & No. of  & No. of & $G$
 & $\epsilon_{\rm pat}$ & $\epsilon_{\rm rig}$
 & $\epsilon_{\rm sel}$ & $\epsilon_{\rm det}$ & $\epsilon_{\rm air}$
 & TOA $\bar{p}$ flux & TOA $\bar{p}/p$ ratio\\
(GeV) & detected $\bar{p}$ & atm. $\bar{p}$ & (m$^{2}$sr)
 & & & & & & (m$^{-2}$s$^{-1}$sr$^{-1}$GeV$^{-1}$) & \\
\hline
0.175 - 0.3 &  3 & 0.17 & 0.30 & 0.73 & 0.96 & 0.68 & 0.49 & 0.87
 & $1.36^{+1.41\,+0.30}_{-0.79\,-0.30} \times 10^{-2}$
 & $0.78^{+0.81\,+0.18}_{-0.45\,-0.18} \times 10^{-5}$\\
0.3   - 0.5 &  7 & 0.78 & 0.31 & 0.73 & 0.92 & 0.74 & 0.62 & 0.88
 & $1.36^{+0.83\,+0.23}_{-0.57\,-0.23} \times 10^{-2}$
 & $0.74^{+0.45\,+0.12}_{-0.31\,-0.12} \times 10^{-5}$\\
0.5   - 0.7 &  7 & 1.4  & 0.31 & 0.73 & 0.87 & 0.74 & 0.65 & 0.89
 & $1.22^{+0.81\,+0.20}_{-0.55\,-0.20} \times 10^{-2}$
 & $0.77^{+0.52\,+0.12}_{-0.35\,-0.12} \times 10^{-5}$\\
0.7   - 1.0 & 11 & 2.8  & 0.32 & 0.72 & 0.80 & 0.73 & 0.67 & 0.89
 & $1.25^{+0.67\,+0.22}_{-0.50\,-0.22} \times 10^{-2}$
 & $1.01^{+0.54\,+0.17}_{-0.40\,-0.17} \times 10^{-5}$\\
1.0   - 1.4 & 15 & 3.5  & 0.32 & 0.72 & 0.72 & 0.56 & 0.68 & 0.90
 & $1.85^{+0.79\,+0.31}_{-0.61\,-0.31} \times 10^{-2}$
 & $1.99^{+0.85\,+0.32}_{-0.66\,-0.32} \times 10^{-5}$\\
\end{tabular}
\end{table}

\narrowtext
Based on these 43 events,
 we obtain the $\bar{p}$ energy spectrum at the top of the atmosphere (TOA)
 in the following way:
The TOA energy of each event is calculated
 by tracing back the particle through the detector material and air
 by using Bethe-Heitler routine of {\sc geant} 3.21.
The corrections are usually small, 30 $\pm$ 3 MeV for a 1 GeV event.
For the particular lowest-energy event which has 157 MeV
 at the center of the detector, the energy correction amounts to 66 MeV
 with estimated accuracy of $\pm$ 2 MeV.
Among the factors necessary to obtain the flux,
 the geometrical acceptance ($G$) can be calculated reliably
 both analytically and by Monte Carlo methods
 due to the simple geometry and uniform magnetic field.
The efficiencies of the pattern-selection ($\epsilon_{\rm pat}$)
 and of the rigidity-selection ($\epsilon_{\rm rig}$)
 for the second-level trigger as well as the off-line selection efficiency
 for the protons are directly determined
 by using the unbiased trigger sample.
The selection efficiency for the $\bar{p}$'s ($\epsilon_{\rm sel}$)
 is obtained by multiplying the proton selection efficiency
 by the ratio (ranging from 1.30 at 0.3 GeV to 1.12 at 1.4 GeV)
 of Monte Carlo simulated selection efficiencies
 for the $\bar{p}$'s to protons.
In this way we minimize the effect of inaccuracy of the simulation.
The {\sc geant/gheisha} code is used for the simulation
 by incorporating the detailed material distribution
 and realistic detector performance in order to also evaluate
 the interaction loss in the instrument ($\epsilon_{\rm det}$).
Care is taken \cite{MATSU,MO97} to use the correct
 $\bar{p}$-nuclei inelastic (including the annihilation) cross sections,
 since original {\sc gheisha} uses cross sections
 much higher than recent measurements, especially below 1 GeV.
The survival probability ($\epsilon_{\rm air}$) of the $\bar{p}$'s
 through the air is estimated by the Monte Carlo simulation.
We also subtract the expected number of atmospheric  $\bar{p}$'s,
 produced by the collisions of cosmic rays in the air.
Among recent calculations \cite{MITSUI,PF96,ST96}
 of the atmospheric $\bar{p}$'s,
 which agree within $\pm$ 20 \% relative accuracy,
 we take the one \cite{MITSUI}
 which utilizes the most detailed nuclear model
 and the 3-dimensional Monte Carlo simulation.

Table \ref{tab:pbsummary} summarizes all relevant quantities
 and resultant $\bar{p}$ fluxes in five energy bins.
The first and the second errors, respectively,
 represent the statistical and systematic errors.
The statistical errors are calculated properly \cite{PART}
 for the Poisson distributions.
The dominant systematic error at low-energy bins is the uncertainty
 in the interaction losses, to which we attribute $\pm$ 40 \% relative error.
At high energy bins,
 the uncertainty in the atmospheric $\bar{p}$ calculations,
 to which we attribute $\pm$ 30 \% relative error,
 becomes an important systematic error.
In all bins, the systematic errors are overwhelmed by the statistical errors.
Table \ref{tab:pbsummary} also contains the $\bar{p}/p$ ratios,
 obtained by counting the number of protons
 which survive the same selections as for the $\bar{p}$'s.
The interaction losses of the protons in the air and in the instrument
 are corrected by the same Monte Carlo simulations,
 and the atmospheric protons are subtracted 
 based on Ref.\cite{PAPINI}.

\begin{figure}[b]
\centerline{\epsfxsize=8.5cm \epsffile{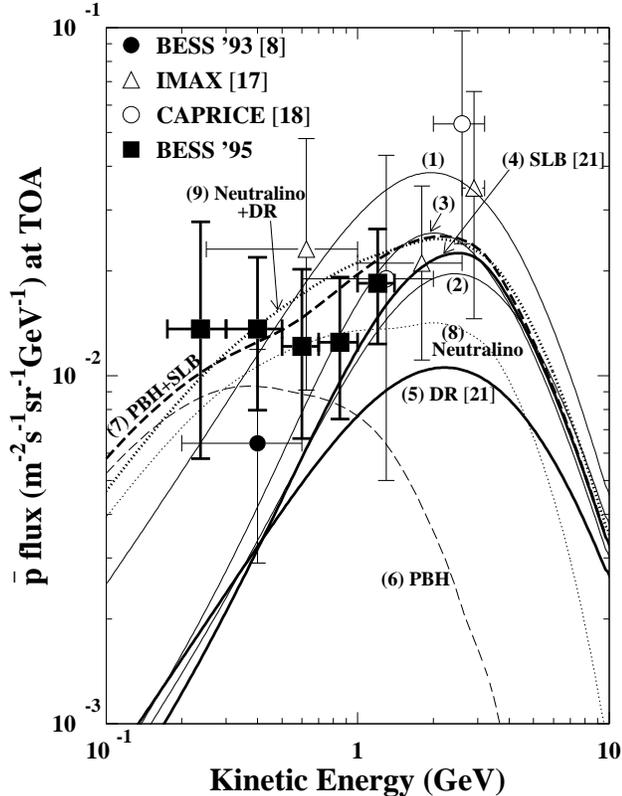}}
\caption{
Comparison of the BESS '95 $\bar{p}$ fluxes with previous data
 and various calculations.
The upper (1) and the lower (2) thin lines represent respectively
 bounds in standard leaky box calculation by Simon {\it et al.} [19].
The middle thin line (3) is the median flux
 calculated in the standard leaky box model by Gaisser and Schaefer [20],
 who attribute a factor 2 ambiguities in both directions.
The upper (4) and the lower (5) thick lines are the calculations
 by Mitsui {\it et al.} [21] in the standard leaky box model
 and in the diffusive reacceleration model (DR) [22], respectively.
The thin (6) and thick (7) dashed lines [23] are, respectively, the spectrum
 from evaporating primordial black holes (with local explosion rate
 of $6 \times 10^{-3}$ pc$^{-3}$yr$^{-1}$) and its sum
 with the standard leaky box model calculation by Mitsui {\it et al}.
The thin (8) and thick (9) dotted lines, respectively, are the spectrum
 from the annihilations of neutralino dark matter [4] and its sum
 with the diffusive reacceleration calculation by Mitsui {\it et al}.
All theoretical fluxes are modulated [24]
 with $\phi_{F}=530$ MV, which corresponds to the BESS '95 flight.
}
\label{fig:pbarflux}
\end{figure}

Our data on the $\bar{p}$ flux are shown in Fig.\ref{fig:pbarflux}
 together with previous measurements and various theoretical calculations.
Our data are consistent with recent measurements (including BESS '93 result),
 all of which have much larger statistical errors.
The energy spectrum we have obtained appears to be flat below 1 GeV
 within present statistical accuracy, and does not exhibit the steep decline
 which seems generic to calculations \cite{SI98,GS92,MI98}
 for the secondary $\bar{p}$'s.
This might be due to a statistical fluctuation,
 might indicate that the propagation models need to be modified,
 or might perhaps suggest a contribution of primary $\bar{p}$ component
 with energy spectrum much softer than the secondary $\bar{p}$.
We note \cite{OR98} that an admixture of $\bar{p}$'s \cite{MI96}
 from evaporating primordial black holes (PBH) \cite{HA75},
 added to a standard leaky box (SLB) calculation \cite{MI98}
 of the secondary $\bar{p}$'s, could provide a fit
 (thick dashed curve in Fig.\ref{fig:pbarflux}) to the data.
Primary $\bar{p}$'s component
 from the annihilation of neutralino dark matter could
 also explain \cite{MI96,OR98} the observed spectrum,
 if we assume an astrophysical enhancement factor of 15 or more,
 which might be possible \cite{MI96}
 due to the clumpiness of the dark matter.
However, given the present large statistical errors and the ambiguities
 in the flux of the secondary $\bar{p}$'s,
\vspace{5cm}
 we cannot exclude the possibility
 of the secondary $\bar{p}$'s alone explaining the data,
 especially if the actual secondary $\bar{p}$ flux
 is among the largest predicted.
To clarify the situation, one needs;
 a) to improve the statistical accuracy,
 b) to detect the predicted peak of the secondary $\bar{p}$'s
 by extending the energy range of the identification
 and to accurately measure its absolute flux,
 and c) to measure the change of the energy spectrum with solar activity.
Concerning the point c), it was shown \cite{MI96} that
 a clear signature of a primary $\bar{p}$ component can be obtained
 by observing an enhancement of low-energy $\bar{p}$'s
 at the solar minimum and a sharp decrease afterward.
Independent to the search for the primary $\bar{p}$ component,
 the precise measurement of the secondary $\bar{p}$ spectrum itself
 will be of crucial importance to determine the propagation mechanism
 of cosmic rays in the Galaxy.
We intend to accomplish these purposes in future BESS flights.

Sincere thanks are given to NASA and NSBF
 for balloon launch and various related support.
We thank Y.~Ajima, M.~Fujikawa, Y.~Higashi, N.~Kimura, T.~Maeno, Y.~Nishihara,
 M.~Otoba, M.~Sasaki and D.~Righter for their help.
The analysis was performed
 by using the computing facilities at ICEPP, Univ. of Tokyo.
This experiment was supported by NASA in the USA,
 and by a Grant-in-Aid for Scientific Research from the Ministry of Education
 and by a Sumitomo Research Grant in Japan.


\begin{references}
\bibitem{RIGIDITY}
 The magnetic-rigidity ($R$ also denoted as rigidity) is defined as $R=P/Z$,
 where $P$ and $Z$ are respectively particle's momentum and charge.
 $R$ is measured from the bending radius in the magnetic field.
\end{references}
\end{document}